\documentclass[12pt]{article}
\usepackage[utf8]{inputenc}

\usepackage{amsmath,amsfonts,graphicx}
\usepackage[square,numbers,sort&compress]{natbib}
\usepackage{fullpage}
\usepackage{lineno}
\usepackage{xcolor}
\usepackage{pdflscape}
\usepackage{hyperref}
\usepackage{multicol,multirow}
\usepackage{authblk}

\newcommand{\vect}[1]{\boldsymbol{\mathbf{#1}}}

\newcommand{\highlight}{}

\title{Estimating excess mortality during the Covid-19 pandemic in Aotearoa New Zealand }

\author[1,*]{Michael J. Plank}
\author[1,2]{Pubudu Senanayake}
\author[3]{Richard Lyon}

\affil[1]{School of Mathematics and Statistics, University of Canterbury, Christchurch, New Zealand}
\affil[2]{Stats NZ, Christchurch, New Zealand}
\affil[3]{Australian Actuaries Institute Mortality Working Group\vspace{0.5cm}}

\affil[*]{Corresponding author: Michael J. Plank, michael.plank@canterbury.ac.nz}

\date{}

\begin{document}


\maketitle

{\bf  Please note an Addendum to this article containing updated results was published in December 2025. This may be found at \url{https://arxiv.org/abs/2512.02266}. }

{\bf The abstract for the addendum is immediately below. The original article and its abstract are on the following pages.}

{\bf Addendum Abstract:}

In our previous article, we estimated excess mortality during in Aotearoa New Zealand for 2020 to 2023. Since our work was published, updated population estimates have been released by Statistics NZ. In this short letter, we provide the results of applying our original model to the new population data. Our updated excess mortality estimate of 2.0\% (95\% CI [0.5\%, 3.3\%]) is 1.3 percentage points higher than our original estimate because the new population estimates for the period 2020 to 2023 are smaller, but the main conclusions of our original article still apply.

\clearpage

\begin{abstract}
{\bf Background.} The excess mortality rate in Aotearoa New Zealand during the Covid-19 pandemic is frequently estimated to be among the lowest in the world. However, to facilitate international comparisons, many of the methods that have been used to estimate excess mortality do not use age-stratified data on deaths and population size, which may compromise their accuracy.

{\bf Methods.} We used a quasi-Poisson regression model for monthly all-cause deaths among New Zealand residents, controlling for age, sex and seasonality. We fitted the model to deaths data for 2014-19. We estimated monthly excess mortality for 2020-23 as the difference between actual deaths and projected deaths according to the model. We conducted sensitivity analysis on the length of the pre-pandemic period used to fit the model. We benchmarked our results against a simple linear regression on the standardised annual mortality rate.

{\bf Results.} We estimated cumulative excess mortality in New Zealand in 2020-23 was 1040 (95\% confidence interval [-1134, 2927]), equivalent to 0.7\% [-0.8\%, 2.0\%] of expected mortality. Excess mortality was negative in 2020-21. The magnitude, timing and age-distribution of the positive excess mortality in 2022-23 were closely matched with confirmed Covid-19 deaths.

{\bf Conclusions.} Negative excess mortality in 2020-21 reflects very low levels of Covid-19 and major reductions in seasonal respiratory diseases during this period. In 2022-23, Covid-19 deaths were the main contributor to excess mortality and there was little or no net non-Covid-19 excess. Overall, New Zealand experienced one of the lowest rates of pandemic excess mortality in the world. 

\end{abstract}

Keywords: age-standardised mortality rate; excess deaths; generalised linear model; Poisson regression; SARS-CoV-2.

\newpage
\section*{Key messages}
\begin{itemize}
    \item {\bf Research question:} Accounting for New Zealand's ageing population, how many additional deaths occurred in 2020-23 relative to what would have been expected in the absence of a pandemic?
    \item {\bf Findings:} Excess mortality in New Zealand in 2020-23 was estimated to be 1040, with a 95\% confidence interval of [-1134, 2927].
    \item {\bf Importance:} New Zealand experienced one of the lowest rates of pandemic excess mortality in the world.
\end{itemize}

\newpage

\section{Introduction}

{\highlight The Covid-19 pandemic caused an estimated 14.8 million deaths globally in 2020 and 2021 \cite{msemburi2023estimates}} Aotearoa New Zealand used a combination of border and community control measures to minimise transmission of SARS-CoV-2 until high vaccine coverage was achieved in late 2021 \cite{baker2020successful,vattiato2022assessment}. 
Following establishment of the B.1.1.529 (Omicron) variant in January 2022 \cite{douglas2022tracing}, New Zealand experienced a series of Covid-19 waves \cite{datta2024impact}. By 31 December 2023, there had been a total of 750 Covid-19-attributed deaths per million people \cite{two_covid_trends}, one of the lowest rates in the world. 

The number of Covid-19-attributed deaths is one measure of the pandemic's impact on mortality, but has limitations. It could be influenced by factors such as test availability, how Covid-19-attributed deaths are distinguished from incidental deaths in Covid-19 positive individuals, and processes for recording multiple causes of death. Also, Covid-19-attributed deaths only measure the direct impact of the disease on mortality. This ignores indirect effects, such as deaths that occurred or were prevented as a result of pandemic control measures, and deaths that would have occurred due to another cause within a given time period. 

Another measure of the pandemic's impact is excess mortality, i.e. the number of additional deaths relative to the expected number if no pandemic had occurred \cite{msemburi2023estimates}. Various methods have been used to estimate the expected number of deaths, known as the baseline. The method of Karlinsky and Kobak \cite{karlinsky2021tracking}, which has been widely used to make international comparisons of excess mortality via the Our World in Data dashboard \cite{HMD,owidcoronavirus}, estimated cumulative excess mortality in New Zealand in 2020-23 as 99 per million. Another widely used model developed by the Economist estimated 198 per million \cite{economist}, whilst Kung et al. \cite{kung2023New} estimated excess mortality for 2020--22 as 215 per million.

The Karlinsky and Kobak \cite{karlinsky2021tracking} baseline relies on extrapolating the pre-pandemic trend in raw death counts. This may, in some cases, mainly reflect the trend in total population size if the crude mortality rate (i.e. deaths per capita) does not vary greatly. Gibson \cite{gibson2024cumulative} correctly pointed out that, in such cases, if the trend in total population size abruptly changes, then the Karlinsky and Kobak baseline will be systematically biased. There was a sharp drop in the rate of population growth in New Zealand in 2020 due to reduced international travel. Gibson \cite{gibson2024cumulative} argued that controlling for population size by fitting to the trend in crude mortality rate, rather than raw deaths, led to a lower baseline than that of Karlinsky and Kobak \cite{karlinsky2021tracking} and, therefore, a higher estimate of excess mortality. 

However, the trend in raw deaths may also be influenced by population ageing, which would generally lead to an upward trend in crude mortality rate. For this reason, where age-stratified data are available, it is preferable to control for age when estimating mortality baselines \cite{hood2024cumulative}. This will limit international comparisons to jurisdictions where age-stratified data are available, but will provide more accurate estimates. Neither Gibson \cite{gibson2024cumulative} nor Kung et al. \cite{kung2023New} controlled for age. Reliable estimates of pandemic excess mortality in New Zealand that control for the combined effects of population growth and ageing are lacking. 

Here, we use age- and sex-stratified data on all-cause mortality in New Zealand to estimate excess mortality in 2020-23 relative to the pre-pandemic baseline. We compare the results of two methods: a quasi-Poisson regression model fitted to monthly age-specific death counts; and a linear regression on the age-standardised annual mortality rate. 

Both methods account for changes in population size and age structure, which means they provide more reliable estimates of excess mortality than methods that do not. We use the quasi-Poisson regression model as our primary method as this allows us to disaggregate results to monthly time periods and by age and sex. We compare aggregated yearly results from this method with the standardised mortality rate linear regression to benchmark our results against a simpler model. We then compare the estimated excess to the number of Covid-19-attributed deaths over time and in different subgroups. This enables us to estimate whether there was any net non-Covid-19 excess mortality during this period.

\section{Methods}

We obtained data on monthly all-cause deaths among New Zealand residents and the Stats NZ estimated resident population \cite{infoshare_erp}, stratified by sex and age, up to the end of 2023. We also obtained data on the number of deaths attributed to Covid-19 \cite{two_covid_trends}. See Supplementary Material Sec. 1.1 for further details. 

We used two methods for constructing a mortality baseline from pre-pandemic data: (1) a {\highlight generalised linear} quasi-Poisson regression model for monthly death counts including age, sex and month of the year as predictors, based closely on the model of the UK's Office for National Statistics \cite{ons2024estimating}; (2) a linear regression fitted to the age- and sex-standardised annual mortality rate. {\highlight Quasi-Poisson regression is a natural model for count data whilst allowing for the possibility that the variance may be higher than in a Poisson regression.} See Supplementary Material Sec. 1.2 for further details.

We fitted each model to all-cause mortality data from January 2014 to December 2019. Using a six-year baseline is a compromise between shorter baselines, which are more sensitive to annual fluctuations in death rates, and longer baselines, which are biased by the nonlinear shape of annual mortality rates over time. We conducted a sensitivity analysis with different lengths of baseline ranging from four to 10 years.

We used the fitted baseline to estimate expected deaths from January 2020 to December 2023 if pre-pandemic trends had continued. We calculated excess mortality as the difference between actual deaths and expected deaths.

\section{Results}

\begin{figure}
    \centering
    \includegraphics[width=\linewidth]{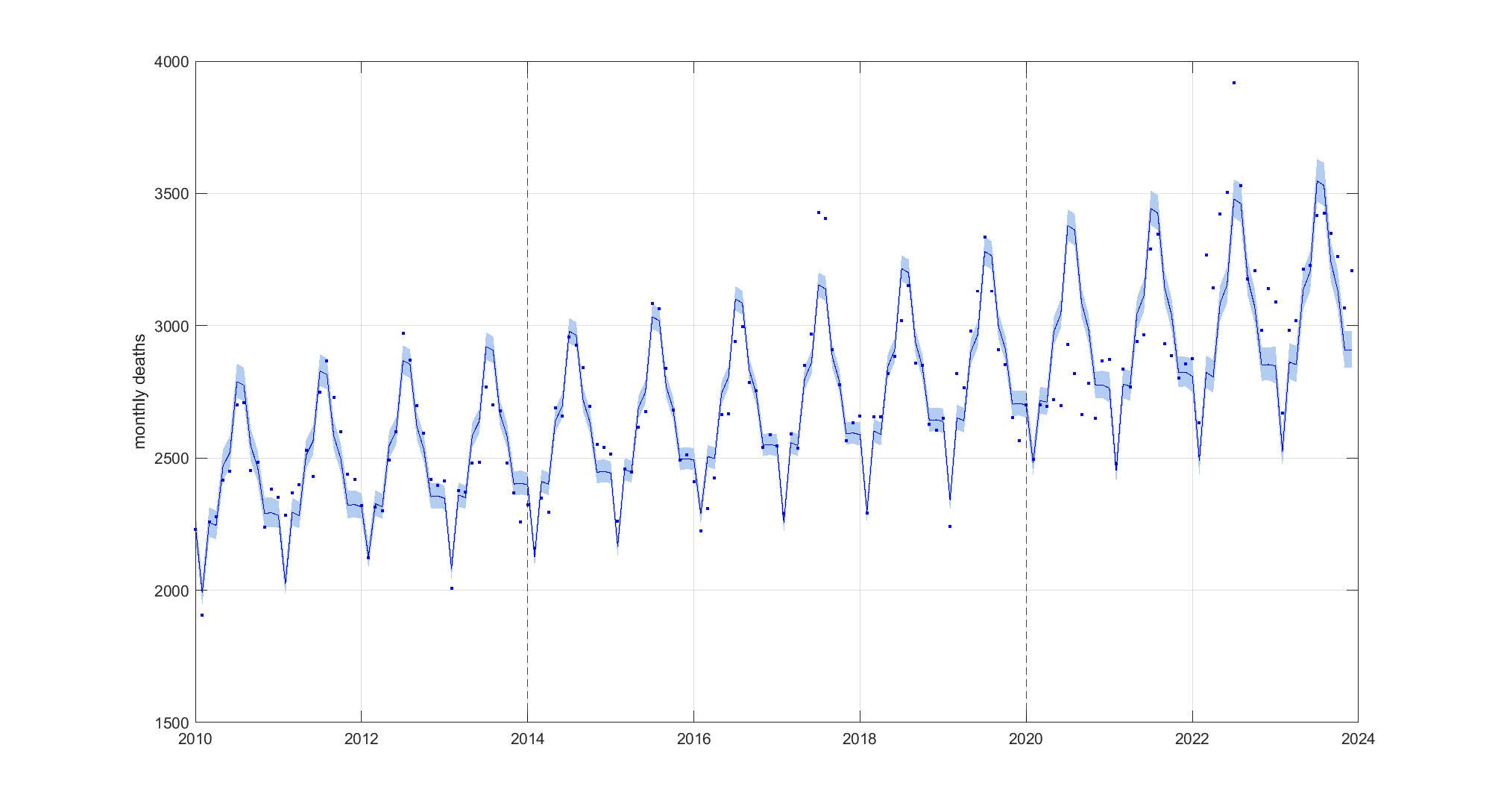}
    \caption{Total monthly all-cause deaths {\highlight in the New Zealand resident population} from January 2010 to December 2023 (blue points) along with the quasi-Poisson regression model fitted to data on monthly deaths from January 2014 to December 2019. Vertical dashed lines show the fitting period. Solid blue curve shows the mean monthly deaths according to the fitted model; shaded band shows the 95\% CI. {\highlight Excess mortality is estimated as the difference between actual deaths for 2020-23 and expected deaths for 2020-23 according to the model (mean and 95\% CI).}\\
    {\bf Alt text:} Graph showing data for total monthly all-cause deaths in the New Zealand resident population from January 2010 to December 2023, alongside the mean and 95\% confidence interval for the quasi-Poisson regression model fitted to the data for the period 2014 to 2019.   }
    \label{fig:total_monthly_deaths}
\end{figure}

\begin{figure}
    \centering
    \includegraphics[width=.6\linewidth]{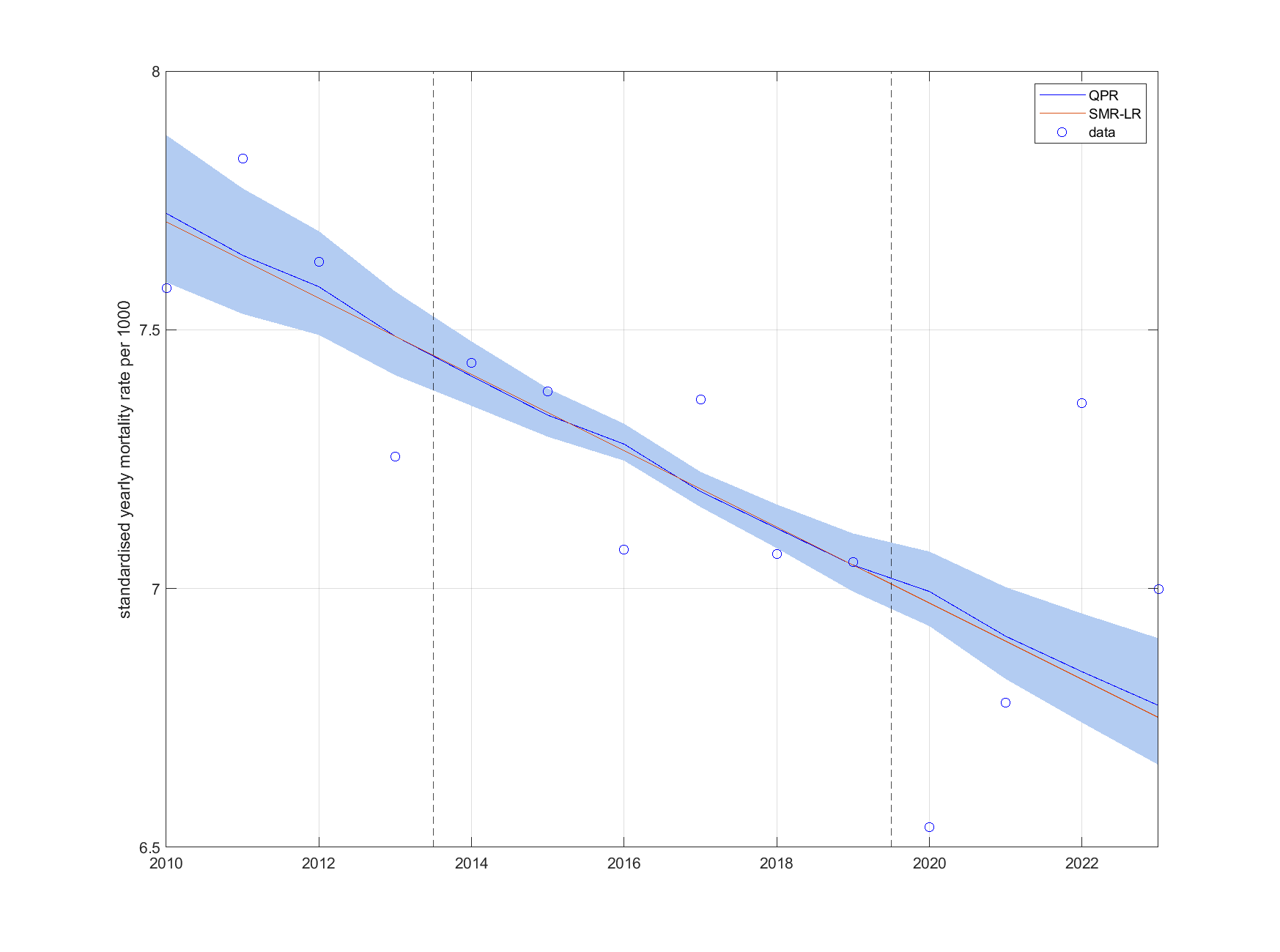}
    \caption{Age- and sex-standardised yearly all-cause mortality rate per 1000 people {\highlight in the New Zealand resident population} (open circles), with the results of the quasi-Poisson regression (QPR) model (solid blue curve = mean, shaded blue band = 95\% CI) and the standardised mortality rate linear regression (SMR-LR) model (solid red). Both models were fitted to data from 1 January 2014 to 31 December 2019 (indicated by the vertical dashed lines). All calculations use the first quarter of 2021 as the standard population.\\
    {\bf Alt text:}  Graph showing data for the age- and sex-standardised yearly all-cause mortality rate per 1000 people in the New Zealand resident population from 2010 to 2023, alongside the mean and 95\% confidence interval for the quasi-Poisson regression model and the standardised mortality rate linear regression model. Both models were fitted to the data for the period 2014 to 2019. }
        \label{fig:smr}
\end{figure}

In the pre-pandemic period, all-cause monthly deaths {\highlight in the New Zealand resident population} had a pronounced seasonal pattern, with relatively high deaths during the winter (i.e. June--August) respiratory disease season (Figure \ref{fig:total_monthly_deaths}). The quasi-Poisson regression model captured the seasonality and the gradually increasing trend in deaths. 
The age- and sex-standardised yearly all-cause mortality rate varied between 7.0 and 7.9 per 1000 between 2010 and 2019, with a decreasing trend (Figure \ref{fig:smr}). The model standardised yearly mortality rate was very similar under the two methods.

\begin{figure}
    \centering
    \includegraphics[width=\linewidth]{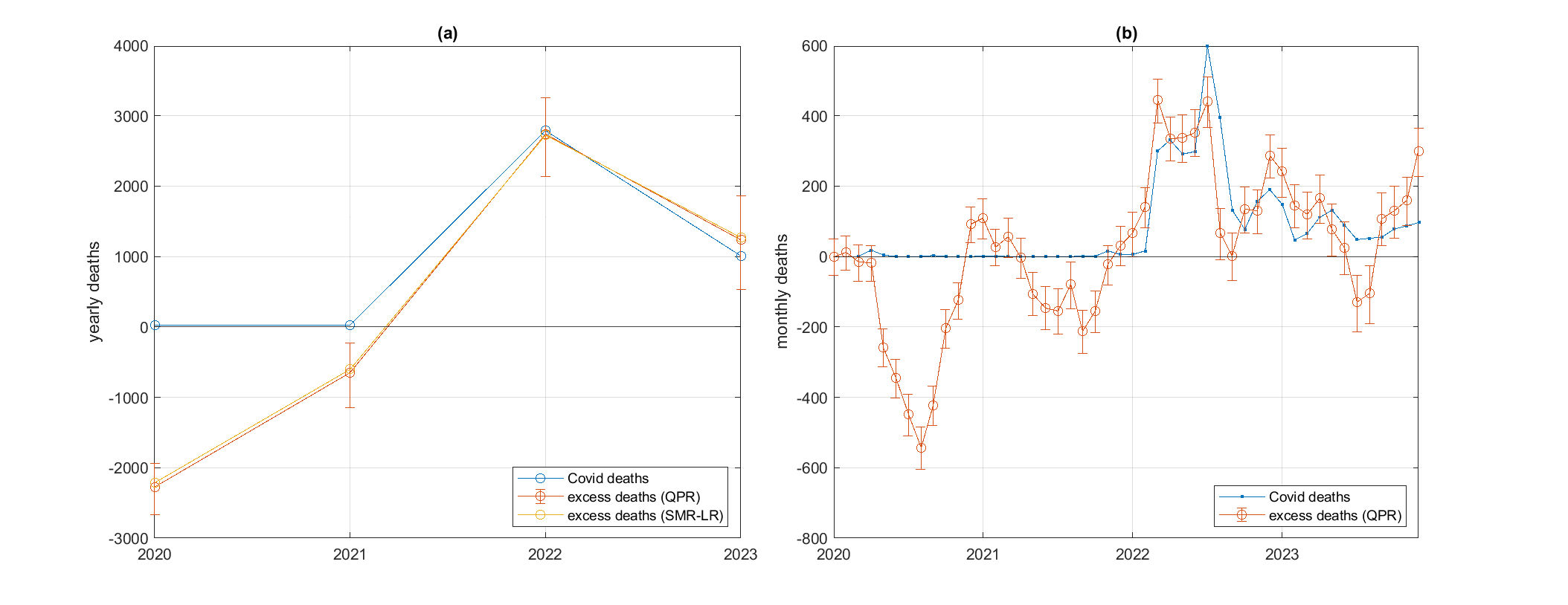}
    \caption{(a) Yearly Covid-19-attributed deaths (blue) along with excess deaths according to the quasi-Poisson regression (QPR) model (red) and the standardised mortality rate linear regression (SMR-LR) model (yellow). (b) Monthly Covid-19-attributed deaths (blue) along with excess deaths according to the QPR model (red). Error bars show the 95\% CI for the QPR model.\\
    {\bf Alt text:} First graph shows the yearly number of Covid-19-attributed deaths from 2020 to 2023 alongside the estimated number of excess deaths according to both the quasi-Poisson regression model and the standardised mortality rate linear regression model. Second graph shows the monthly number of Covid-19-attributed deaths from January 2020 to December 2023 alongside the estimated number of excess deaths according the quasi-Poisson regression model.    }
    \label{fig:estimated_excess}
\end{figure}

The number of excess deaths according to the quasi-Poisson regression model was -2276 (95\% CI [-2663, -1941]) in 2020; -654 [-1144, -228] in 2021; 2735 [2141, 3252] in 2022; and 1235 [531, 1858] in 2023 (Figure \ref{fig:estimated_excess}a, red). The cumulative excess for 2020-23 was 1040 [-1134, 2927], which equates to 204 [-222, 573] per million and to 0.7\% [-0.8\%, 2.0\%] of expected mortality. The standardised mortality rate linear regression model gave very similar results (Figure \ref{fig:estimated_excess}a, yellow). 

Covid-19-attributed deaths were extremely low in 2020 and 2021 (26 in each year). 
In 2022-23, the number of Covid-19-attributed deaths was close to the central estimate for excess deaths (Figure \ref{fig:estimated_excess}a, blue), accounting for 102\% of the estimated excess in 2022 and 82\% in 2023, and fell within the 95\% CI in both years. 

Looking at monthly data (Figure \ref{fig:estimated_excess}b), Covid-19 deaths and excess deaths according to the quasi-Poisson regression model followed similar trends in 2022-23. The number of Covid-19 deaths was below the 95\% CI for excess deaths in February and March 2022, and above it from July to September 2022. The unexplained excess in early 2022 could represent unattributed Covid-19 deaths that occurred when the country's laboratory testing capacity became overwhelmed \cite{moh2022covid}. The monthly discrepancies could also partly reflect temporal displacement of mortality due to disruption of seasonal respiratory disease patterns \cite{huang2024impact}. 

\begin{figure}
    \centering
    \includegraphics[width=.8\linewidth]{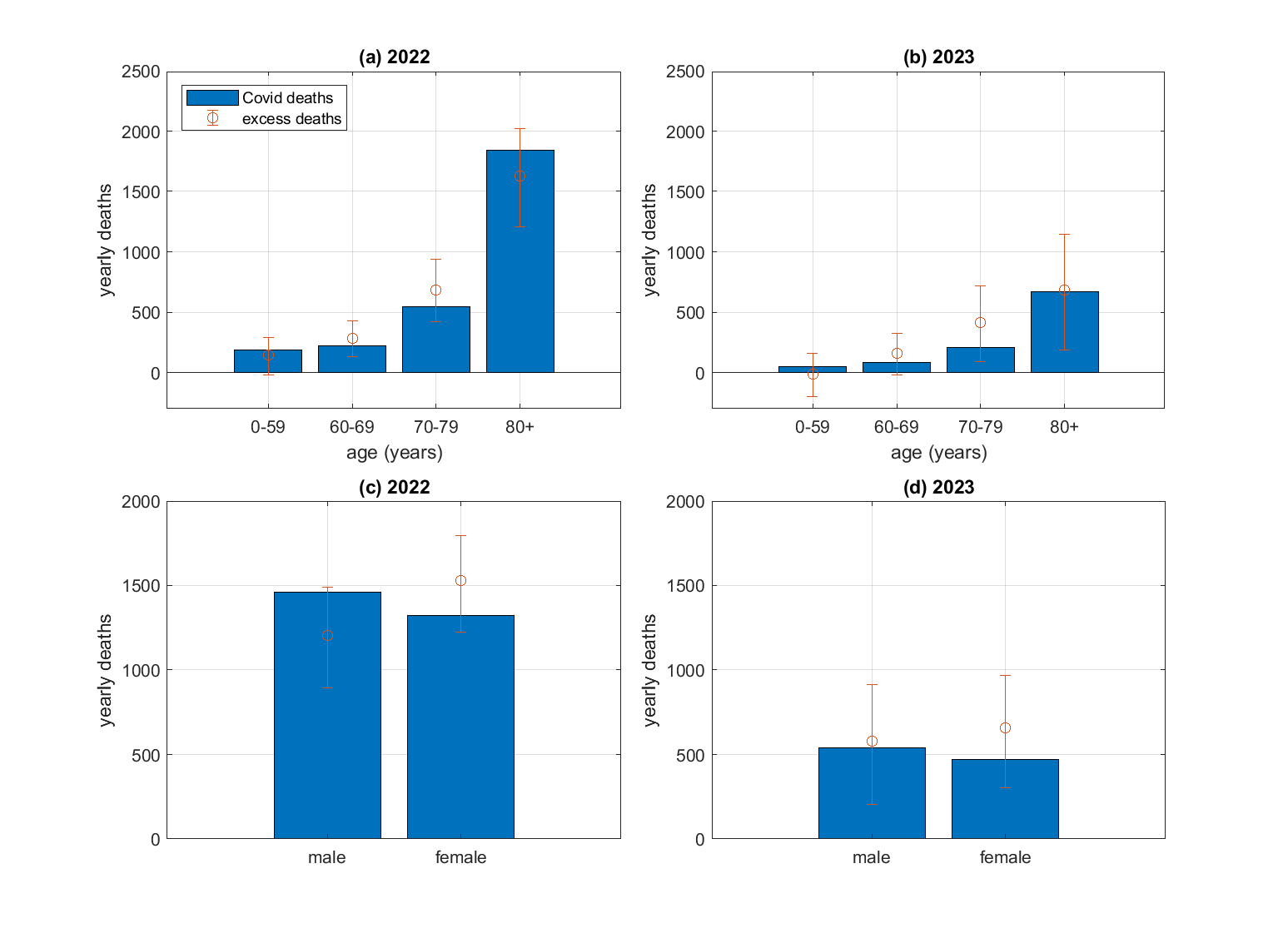}
    \caption{Yearly Covid-19-attributed deaths (blue) and excess deaths (mean and 95\% CI) according to the quasi-Poisson model (red) disaggregated by age (a,b) or sex (c,d) in 2022 and 2023.\\
    {\bf Alt text:} Bar charts showing the yearly number of Covid-19 attributed deaths alongside the estimated number of excess deaths according to the quasi-Poisson regression model. The top two graphs show the numbers of deaths in the 0-59, 60-69, 70-79 and 80+ age groups for 2022 and 2023 respectively. The bottom two graphs show the number of deaths among male and female individuals in 2022 and 2023 respectively.    }
    \label{fig:deaths_by_age}
\end{figure}

The number of Covid-19 deaths fell within the 95\% CI for the number of excess deaths in each age band (0-59 years, 60-69 years, 70-79 years and 80+ years) and for both sexes in 2022 and 2023 (Figure \ref{fig:deaths_by_age}). Note we used these four age bands because appropriate data on Covid-19 deaths were available for comparison (see Supplementary Figure S1 for results disaggregated into finer age groups).
The results in Figure \ref{fig:deaths_by_age} mirror similar patterns seen in Australia \cite{mortality2024how}, where there was some positive non-Covid excess in the 70-79-year age band and, despite there being more Covid-19 deaths among men than women, the central estimate for excess deaths was higher for women.

When different pre-pandemic fitting periods were used, the estimated number of excess deaths for 2020-23 according to the quasi-Poisson regression model varied between a low of -1770 [-4973, 924] for the four-year baseline, and a high of 1995 [500, 3368] for the nine-year baseline (Table \ref{tab:sensitivity}). Expressed as a percentage of expected deaths, these correspond to a low of -1.2\% [-3.4\%, 0.6\%] and a high of 1.4\% [0.4\%, 2.4\%]. The estimate from our default baseline of six years (2014-19) ranked third out of the seven baselines investigated. The 95\% CI for the quasi-Poisson model contained zero in all cases except for the nine-year baseline.   The results of the standardised mortality rate linear regression model showed a similar pattern with respect to baseline length. {\highlight In both models,  the variations in estimated excess seen in Table  \ref{tab:sensitivity} were mainly driven by whether or not the first year in the fitting period had a relatively high all-cause death rate. }

The Karlinksy and Kobak estimate for 2020-23 of 517 excess deaths  \cite{karlinsky2021tracking,owidcoronavirus} falls within the 95\% CI for all seven baselines. The Economist estimate of 1026 excess deaths \cite{economist} is very close to our central estimate of 1040 for the six-year baseline, is within the 95\% CI for five of the seven baselines, and slightly above it for other two. These estimates are therefore largely consistent with our results. 
In contrast, Gibson estimated $+4\%$ excess mortality for the three-year period of 2020-22 using a five-year baseline, which translates to approximately 4000 excess deaths \cite{gibson2024cumulative}. This estimate, which failed to control for changes in population age structure, is 4500 deaths higher than our central estimate for this period using a five-year baseline, and 2500 deaths higher than the highest upper limit of the 95\% CI for {\em any} baseline length between four and 10 years (see Supplementary Table S1).

\begin{table}
\centering
\begin{tabular}{lrrrr} 
\hline
Baseline period & \multicolumn{3}{l}{Excess (QPR)} & Excess (SMR-LR) \\ 
\hline
2016--2019    &    -1770 [ & -4973, & 924]   &   -1489 \\ 
2015--2019    &    456 [ & -1999, & 2743]   &   641 \\ 
2014--2019    &    1040 [ & -1134, & 2927]   &   1177 \\ 
2013--2019    &    -784 [ & -2687, & 963]   &   -948 \\ 
2012--2019    &    510 [ & -1152, & 2018]   &   611 \\ 
2011--2019    &    1995 [ & 500, & 3368]   &   2271 \\ 
2010--2019    &    1148 [ & -246, & 2402]   &   1258 \\ 
\hline
\end{tabular}

\caption{Sensitivity analysis showing the cumulative estimated number of excess deaths for the four-year period 2020-23 with different length baselines. In each row of the table, the models were fitted to deaths data in a period of between four and 10 years ending in 2019. Results are shown for the quasi-Poisson regression (QPR) model (mean and 95\% CI) and the standardised mortality rate linear regression (SMR-LR) model. }
\label{tab:sensitivity}
\end{table}

\section{Discussion}

We have estimated the number of excess deaths in Aotearoa New Zealand between January 2020 and December 2023 using two different methods. Importantly, both methods controlled for age, sex and population size (one by including these variables as predictors in a regression model, and one by using an age- and sex-standardised mortality rate). This avoids the potential for systematic {\highlight biases} caused by changes in population size or age structure. 

To comprehensively measure the impact of the pandemic on deaths, estimates need to include the period after most non-pharmaceutical interventions were removed and the virus became endemic. However, the longer pre-pandemic baselines are extrapolated forward in time, the greater the uncertainty and potential for bias due to nonlinear trends over time. In New Zealand, most remaining interventions ended in September 2022 \cite{nzgovt2022covid}, except for mask requirements in healthcare and aged residential care settings and isolation requirements for confirmed cases, which continued until August 2023 \cite{nzgovt2023all}. Choosing the period from January 2020 to December 2023 is a reasonable trade-off that allows enough time for the transmission dynamics to settle into a relatively steady pattern without extrapolating pre-pandemic trends too far. Organisations estimating excess deaths in the UK \cite{ons2024estimating} and Australia \cite{mortality2024how} have adopted a similar approach, moving to a new baseline from 2024 that partly includes the pandemic period.

Consistent with other methods \cite{karlinsky2021tracking,economist}, both our models estimated negative excess mortality in 2020 and, to a lesser extent, 2021. This was likely due to the effective suppression of Covid-19 (there were only 52 Covid-19 deaths in 2020-21), the elimination of influenza \cite{cheung2024severe}, and the reduction in other respiratory pathogens during this period \cite{turner2024comparison,huang2024impact}.

Our models estimated positive excess mortality in 2022 and 2023, after New Zealand ended its elimination strategy and the Omicron variant became established. The magnitude, timing and age-distribution of the excess closely matched those of Covid-19-attributed deaths. 
This suggests that the bulk of the excess in 2022-23 was directly attributable to Covid-19, either as an underlying or a contributory cause of death. This could be because there were relatively few Covid-19 deaths that were not recorded as such. Alternatively it could be that any undocumented Covid-19 deaths were offset by reduced mortality from other causes, such as non-Covid respiratory disease. 

The timing of the excess does not coincide with the timing of the Covid-19 vaccine rollout, the bulk of which occurred between February 2021 and February 2022 (more than 10 million doses were administered in this period). Excess deaths were either negative or close to zero for most of this period. In contrast, the period during which excess deaths were relatively high (March 2022 to April 2023) coincided with a period where far fewer vaccine doses (fewer than 1.7 million) were administered. This suggests that vaccines were not responsible for any substantial portion of the excess mortality, consistent with New Zealand safety data \cite{medsafe_adverse} and international research showing very low risk of severe adverse events following Covid-19 vaccination \cite{cheung2024severe}. 

Mortality rates fluctuate from year to year depending on the severity of the winter respiratory illness season and other factors. This means that the fitted baseline for expected mortality can be sensitive to the choice of fitting period. To mitigate this, we ran our model with seven different fitting periods ranging from four to 10 years duration. The limits of the 95\% CIs for excess mortality for 2020-23 fell between -3.4\% and +2.4\% and contained zero for all but one choice of fitting period. Thus our results provide robust evidence that, regardless of the choice of baseline, there was no more than 2.4\% cumulative excess mortality between 2020 and 2023 and any net excess that did occur cannot be confidently distinguished from zero. 

We have not attempted to apply our method to estimate excess mortality in countries other than New Zealand, as this is beyond the scope of the study. Nevertheless, our estimated excess mortality of 204 [-222, 573] deaths per million people for 2020-23 ranks amongst the lowest in the world. Estimated excess mortality for 2020-23 for most OECD countries (with a few notable exceptions including Australia) was in the range 1500--4500 per million \cite{karlinsky2021tracking,HMD,owidcoronavirus}.

Gibson \cite{gibson2024cumulative} argued that, by extrapolating raw death counts and not accounting for the abrupt drop in population growth in 2020, the Karlinksy and Kobak method \cite{karlinsky2021tracking} systematically overestimated expected mortality, and therefore underestimated excess deaths.  {\highlight However, although population growth slowed dramatically during the pandemic, this was largely a result of reduced migration, which occurs predominantly in younger groups \cite{infoshare_migration}. The population over 65 years of age (which is only around 15\% of the total population but accounts for 80\% of deaths) continued to grow at a rate similar to before the pandemic (see Supplementary Figure S2). By controlling for total population size but not for age, Gibson's method introduced a systematic bias and consequently underestimated expected mortality for 2020-22 by approximately 4\%. }
The main advantage of the method of Karlinsky and Kobak \cite{karlinsky2021tracking} is that it can be applied to a large set of countries to provide a meaningful and useful basis for international comparisons. Despite {\highlight not controlling for population size or age}, the Karlinksy and Kobak estimate was within the 95\% CIs from our model. {\highlight This is because the pre-pandemic trend in death counts was driven more by population ageing (which continued throughout the pandemic) than by population growth (which did not).}

Our study has a number of important limitations. 
Due to reporting delays, there could be approximately 40 deaths that occurred in 2023 that do not yet appear in the data. Hence, our results could underestimate excess deaths for 2023 by around 40, but we note this difference is much smaller than the model confidence intervals.

We have compared estimates of all-cause excess mortality with Covid-19-attributed deaths. It is likely that the pandemic differentially impacted different causes of death at different times. This could be investigated by estimating excess mortality associated with different causes of death \cite{mortality2024how}. However, we have not attempted this due to lack of up-to-date data on causes of death other than Covid-19. 

Our model lumped all deaths over 95 years old into a single age class. This was necessary because population size data was only available at this level. The size of the over 95 year age group increased from 5130 in 2014-Q1 to 8530 in 2023-Q4. It is likely that the age-distribution within this group became increasingly elderly over time, which would influence the group-specific mortality rate. This is mitigated by the inclusion of an interaction term in the quasi-Poisson model, which allows a group-specific linear trend in the mortality rate. Nevertheless, the model cannot explicitly account for any shift in the age distribution within the group that may have occurred between 2019 and 2023.

Covid-19 in New Zealand disproportionately affected M\=aori, Pacific Peoples, and people living with high levels of deprivation \cite{steyn2021maori,whitehead2023inequities,satherley2024individual}. This mirrors well-documented international trends \cite{mathur2021ethnic,nguyen2024racial} whereby social determinants of health intersect with factors affecting people's ability to take protective measures, such as working from home \cite{abrams2020covid}. Our analysis did not include socioeconomic variables such as ethnicity or deprivation because the required data was not available. Investigating how excess mortality in 2022 and 2023 was distributed across these axes, and how this relates to the distribution of Covid-19 deaths, is an important aim for future research.

\subsection*{Ethics approval}
Ethics approval was not required for this research as it only used routinely collected administrative data, which was de-identified and aggregated prior to use.

\subsection*{Data availability}
Data and code to reproduce the analysis are available at \url{https://dx.doi.org/10.5281/zenodo.15107131}.
The results shown in the article used raw, unrounded data on death counts. The raw data cannot be published due to privacy concerns relating to small counts. To preserve confidentiality, the repository linked above contains a data set where death counts were randomly rounded (see Supplementary Material Sec. S1.1 for details). Running the code on this dataset will produce results that are similar to those in the article, but not identical and with broader confidence intervals due to the added noise. 

\subsection*{Author contributions}
MJP: conceptualisation, methodology, software, validation, formal analysis, writing -- original draft, writing -- review and editing. 
PS: conceptualisation, methodology, software, validation, data curation, writing -- review and editing.
RL: methodology, writing -- review and editing.

\subsection*{Use of artificial intelligence (AI) tools}
AI tools were not used in this work.

\subsection*{Funding}
None.

\subsection*{Acknowledgements}
MJP acknowledges support from Te Niwha Infectious Diseases Research Platform, Institute of Environmental Science and Research, grant number TN/P/24/UoC/MP. The authors are grateful to Jennifer Brown and Philipp Wacker for helpful discussions on confidence intervals for quasi Poisson regression models, to Rebekah Hennessey and Helen He for assistance with sourcing the deaths data, and to two anonymous peer reviewers for helpful comments on an earlier version of this manuscript.

\subsection*{Conflict of interest}
None declared.

\clearpage
\setcounter{page}{1}
\setcounter{section}{0}
\setcounter{equation}{0}
\setcounter{figure}{0}
\setcounter{table}{0}
\renewcommand\thepage{S\arabic{page}}
\renewcommand\thesection{S\arabic{section}}
\renewcommand\theequation{S\arabic{equation}}
\renewcommand\thefigure{S\arabic{figure}}
\renewcommand\thetable{S\arabic{table}}

\begin{center}
    {\Large\bf Supplementary Material}
\end{center}

\section{Supplementary methods}

Data and Matlab code to reproduce the results in the article are archived and publicly available \url{https://dx.doi.org/10.5281/zenodo.15107131}.

\subsection{Data}

{\bf All-cause mortality data.} We obtained data from Stats NZ on the monthly number of all-cause deaths by date of death among New Zealand residents between January 2010 and December 2023. The data were stratified by sex and age in one-year age groups. The final age group was an open-ended age group for those 95 years of age and older. 
The data was extracted in November 2024 and Stats NZ consider it to be 99.9\% complete for deaths that occurred in the study period (i.e. up to 31 December 2023). 

The results shown in the article used the raw, unrounded data on death counts. The raw data cannot be published due to privacy concerns relating to small counts. To preserve confidentiality, the repository linked above contains a data set where death counts are randomly rounded as follows:
\begin{itemize}
\item True counts under 6 were randomly rounded to 0 or 6, with the probability of rounding proportional to the distance of the true count to 0 or 6 (e.g., a true count of 5 has a higher chance of being rounded to 6 than to 0).
\item True counts 6 or above were randomly rounded to base 3. This means that counts were randomly rounded up or down to the nearest value divisible by 3, with the probability proportional to the distance (e.g., a true count of 8 may be rounded to 6 or 9, with a higher chance of being rounded to 9).
 \end{itemize}
Running the code on this dataset will produce results that are similar to those in the article, but not identical and with broader confidence intervals due to the added noise.

{\bf Population data.} We used the published Stats NZ estimated resident population (ERP), stratified by sex and age in one-year age groups, quarterly between 1991-Q1 and 2023-Q4 \cite{infoshare_erp}. The final age group was an open-ended age group for those 95 years of age and older. 

The ERP, at any given point in time, is an estimate based on a starting ``base population'' and accounting for subsequently observed births, deaths (natural increase/decrease) and net migration \cite{stats2021population}. This base population is the coverage-adjusted resident population established from the most recent census, its associated post-enumeration survey (PES), and an estimate of residents overseas during the census \cite{stats2020estimated}. 

The base population is updated after each census, in a process termed ``rebasing'' \cite{stats2021population}, with the most recent available being Census 2018. At the time of writing, the Census 2023 based ERP was not yet available. As with every census update, we expect a revision to the population estimates for 2023 (and to a lesser extent 2019-2022) when the new census becomes available. For example, updating from Census 2013 to Census 2018 increased the estimated population for 2018 by 60,000 (approximately 1.2\%). The largest difference was among males aged between 30 and 44 years, although 15--29 year olds, and 45--59 year olds also showed notable differences \cite{stats2020estimatedb}.
These differences were driven primarily by inaccuracies in migration estimates and, to a lesser extent, births and deaths between censuses. They may also have been influenced by inaccuracies in the various census counts and associated coverage adjustments; and potential mismatch between the definition of a New Zealand resident and census respondents' interpretation of this. 

Given that the components of population change are primarily driven by administrative data and now use observed outcomes for migration \cite{stats2024migration}, inaccuracies in the components of population change should be small for the period between Census 2018 and Census 2023. However, the other factors remain pertinent, and therefore it is likely that the estimated population will be revised for 2023 based on the latest census results. 

{\bf Covid-19-attributed mortality data.} We obtained data for the daily number of deaths attributed to Covid-19 between 29 March 2020 (the date of the first recorded Covid-19 death in New Zealand) and 31 December 2023 from Te Whatu Ora (Health New Zealand) Covid-19 Trends and Insights Dashboard \cite{two_covid_trends}. Te Whatu Ora defines Covid-19-attributed deaths as those where Covid-19 was recorded as either the underlying cause of death or a contributory cause of death. Deaths were stratified by coarse age band (under 60 years, 60--69 years, 70--79 years, and 80 years and over) and by sex (but not by age and sex together). We accessed this dataset on 20 November 2024 and it is expected to be complete for deaths that occurred up to 31 December 2023. We aggregated daily death counts into monthly totals.

\subsection{Models}

\subsubsection*{Method 1: Quasi-Poisson regression model}
We fitted a quasi-Poisson generalised linear regression model to death counts $d_{it}\sim\mathrm{QuasiPoisson}(\mu_{it})$ in each stratum $i$, consisting of a combination of sex and one-year age group, and monthly time period $t$. We used a similar model specification to that used by the UK's Office for National Statistics \cite{ons2024estimating} for the expected number of deaths $\mu_{it}$ in stratum $i$ and time period $t$:
\begin{eqnarray} 
\log(\mu_{it}) &=& \beta_0 + \beta_1 t  + \beta_2(\mathrm{age}_i) + \beta_3(\mathrm{sex}_i) + \beta_4(\mathrm{month}_t) + 
\beta_5(\mathrm{ageCoarse}_i\times t) \nonumber \\
&& + \beta_6(\mathrm{ageCoarse}_i\times\mathrm{sex}_i) 
+ \beta_7(\mathrm{ageCoarse}_i\times\mathrm{month}_t) \nonumber \\
&& + \log(N_{it}) + \log(\mathrm{days}_t ),  \label{eq:GLM}
\end{eqnarray}
where $N_{it}$ is the population size in stratum $i$ at time $t$. The time variable $t$ was an integer-valued variable measured in months since January 2014. The variables $\mathrm{age}_i$, $\mathrm{sex}_i$ and month of the year ($\mathrm{month}_t$) were treated as categorical variables, meaning that a separate coefficient $\beta$ was estimated for each value of the categorical variable (except for the baseline value). This meant estimating 95 coefficients for $\mathrm{age}_i$ (one for each age group from 1--2 years up to 95 years and over), 1 coefficient for $\mathrm{sex}_i=\mathrm{female}$ and 11 coefficients for $\mathrm{month}_t$ (representing February to December). 

We included interaction terms for age and time, age and sex, and age and month of the year. To avoid fitting too many coefficients, the interaction terms used a coarse age grouping variable ($\mathrm{ageCoarse}_i$), defined as under 30 years, 30--69 years, and five-year age bands from 70 years up to 95 years and over (giving a total of eight coarse age groups). This meant that $7\times 1=7$ coefficients were estimated for the age:time interaction (recalling that time is a numerical not a categorical variable), $7\times 1=7$ for the age:sex interaction and $7\times 11=77$ for the age:month interaction. This way of modelling interactions follows the UK ONS method and allowed for the trend in mortality rate, the difference in mortality rate between the sexes, and the effect of seasonality to be different in different coarse age bands (but the same in all one-year age groups within the same coarse age band). 

The population size ($N_{it}$) in stratum $i$ at time $t$ and the number of days in the month ($\mathrm{days}_t$) were included as offsets, meaning that the expected number of deaths $\mu_{it}$ in stratum $i$ was assumed to be directly proportional to $N_{it} \mathrm{days}_t$.

The model was fitted to data for the six-year pre-pandemic period January 2014 to December 2019. We investigated alternative fitting periods in a sensitivity analysis.

To account for model uncertainty, we sampled $M=5000$ coefficient vectors $\vect{\beta}$ from its approximate sampling distribution $N( \hat{\vect{\beta}}, \Sigma)$, where $\hat{\vect{\beta}}$ is the maximum likelihood estimate for $\vect{\beta}$ and $\Sigma$ is its covariance. For each coefficient vector $\vect{\beta}^{(j)}$ ($j=1,\ldots,M$), we calculated the mean number of expected deaths $\mu_{it}^{(j)}$ in each stratum $i$ and time period $t$ according to Eq. \eqref{eq:GLM}. We aggregated these predictions across combinations $S$ of strata and time periods (e.g. all age and sex groups in a given month or year, or specific age bands within a given month or year) to produce a set of $M$ samples, $\mu_{S}^{(j)}$, for each output of interest:
\begin{equation} \label{eq:aggregation}
    \mu_{S}^{(j)}=\sum_{(i,t)\in S} \mu_{it}^{(j)}.
\end{equation}
We calculated 95\% confidence intervals (CI) for aggregated outputs $\mu_{S}$ as the range from the 2.5$^\mathrm{th}$ to the 97.5$^\mathrm{th}$ percentile of the $M$ values of $\mu_{S}^{(j)}$. Excess deaths $E_S$ for any combination $S$ of strata and time periods were calculated as the difference between actual deaths and mean expected deaths according to the fitted model:
\begin{equation} \label{eq:excess}
    E_S = d_S - \mu_S.
\end{equation}
We calculated 95\% CIs for $E_S$ in the same way by using the 2.5$^\mathrm{th}$ and 97.5$^\mathrm{th}$ percentile values for $\mu_S$ in Eq. \eqref{eq:excess}.

\subsubsection*{Method 2: standardised mortality rate linear regression}

We calculated the age- and sex-standardised yearly mortality rate $m_Y$ for year $Y$ as follows:
\begin{equation}
    m_Y = \frac{1}{\sum_i N_{i,\mathrm{std}} } \sum_i \frac{  \left(\sum_{t\in Y} d_{it} \right) N_{i,\mathrm{std}} }{ \bar{N}_{i,Y} } , 
\end{equation}
where $N_{i,\mathrm{std}}$ is the size of the standard population in stratum $i$ and $\bar{N}_{i,Y}$ is the mean population in stratum $i$ in year $Y$, calculated by averaging the four quarterly population sizes. We used the first quarter of 2021 as the standard population, though results are very similar for different choices of time period for the standard population. We used the standard population as the denominator when presenting results for excess deaths per million people.

We fitted a simple linear regression to the standardised yearly mortality rate in the six-year period from 2014 to 2019 (with alternative fitting periods investigated in a sensitivity analysis). We calculated the expected standardised mortality rate $\tilde{m}_Y$ for each of the years 2020 to 2023 by extrapolating this linear regression forwards in time. We then calculated the estimated excess deaths $E_Y$ for year $Y$ as
\begin{equation}
    E_Y = \left( m_Y - \tilde{m}_Y\right) \sum_i N_{i,\mathrm{std}}.
\end{equation}

\clearpage

\section{Supplementary results}

\begin{table}[h!]
\centering
\begin{tabular}{lrrrr} 
\hline
Baseline period & \multicolumn{3}{l}{Excess (QPR)} & Excess (SMR-LR) \\ 
\hline
2016--2019    &    -1984 [ & -4080, & -213]   &   -1812 \\ 
2015--2019    &    -569 [ & -2214, & 961]   &   -443 \\ 
2014--2019    &    -195 [ & -1668, & 1081]   &   -93 \\ 
2013--2019    &    -1385 [ & -2695, & -200]   &   -1496 \\ 
2012--2019    &    -526 [ & -1687, & 512]   &   -456 \\ 
2011--2019    &    464 [ & -581, & 1421]   &   660 \\ 
2010--2019    &    -98 [ & -1069, & 769]   &   -26 \\ 
\hline
\end{tabular}
\caption{Total estimated number of excess deaths for the three-year period 2020-22 with different length baselines. In each row of the table, the models were fitted to deaths data in a period of between four and 10 years ending in 2019. Results are shown for the quasi-Poisson regression (QPR) model (mean and 95\% CI) and the standardised mortality rate linear regression (SMR-LR) model. }
\label{tab:sensitivity}
\end{table}

\clearpage

\begin{figure}
    \centering
    \includegraphics[width=\linewidth]{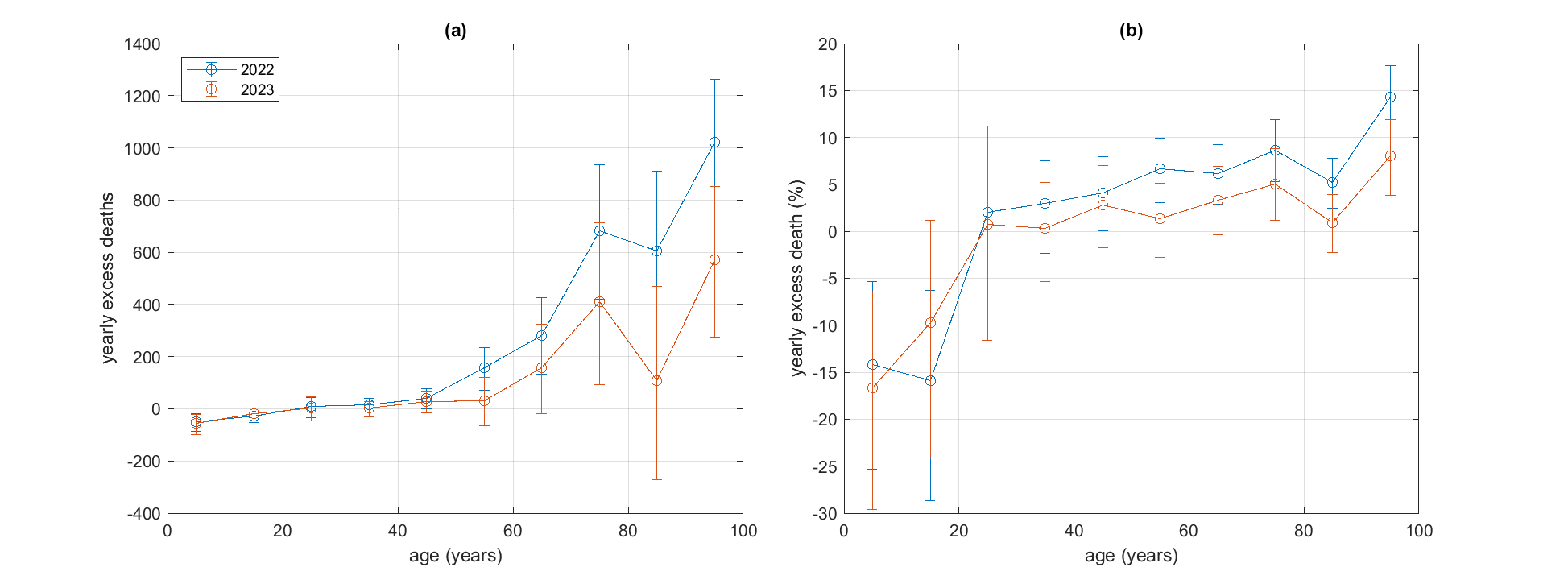}
    \caption{Yearly estimated excess deaths for 2022 (blue) and 2023 (red) in ten-year age bands according to the quasi-Poisson regression model fitted to a six-year baseline: (a) number of excess deaths; (b) excess deaths as a percentage of expected deaths. Points are plotted at the midpoint of their respective ten-year age band and the last point represents over-90-year-olds.  }
    \label{fig:S1}
\end{figure}

\clearpage

\begin{figure}
    \centering
    \includegraphics[width=\linewidth]{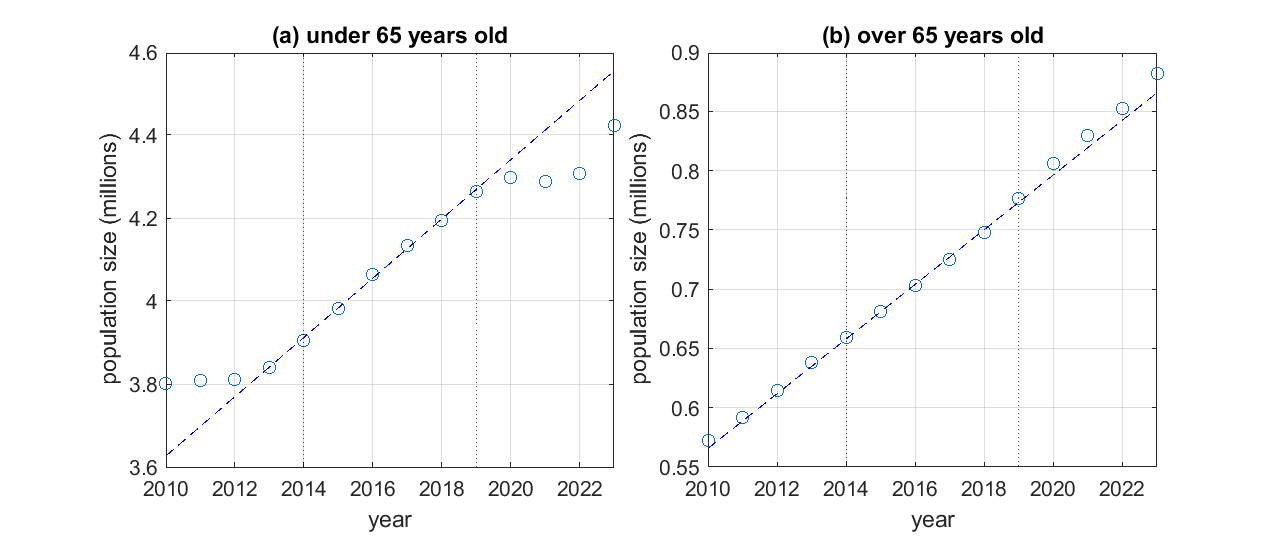}
    \caption{Population size over time for people aged: (a) under 65 years; (b) 65 years and over (blue circles) together with a linear regression (dashed blue line) fitted to data from 2010 to 2019 (fitting period indicated by vertical dotted lines). Comparing the data for 2020 to 2023 with the projected pre-pandemic trend shows that: (a) the growth rate of the population under 65 years old abruptly slowed in 2020; (b) the size of the population over 65 years old continued to grow at a similar rate as in the pre-pandemic period. Over 65-year-olds are approximately 15\% of the total population, but account for approximately 80\% of all deaths. Therefore, a model that uses total population size to project expected deaths without accounting for age will systematically underestimate expected mortality.   }
    \label{fig:S2}
\end{figure}

\clearpage


\end{document}